# Stable fixed points in the Kuramoto model

**Richard Taylor**

F2-01-017, 24 Fairbairn Avenue, Fairbairn Business Park, Canberra ACT Australia 2600

E-mail: richard.taylor@dsto.defence.gov.au

**Abstract**

We develop a necessary condition for the existence of stable fixed points for the general network Kuramoto model, and use it to show that for the complete network the homogeneous model has no non-zero stable fixed point solution. This result provides further evidence that in the homogeneous case the zero fixed point has an attractor set consisting of the entire space minus a set of measure zero, a conjecture of Verwoerd and Mason (2007).

**1 Introduction**

The Kuramoto model (see Kuramoto (1975)) was originally motivated by the phenomenon of collective synchronization whereby a system of oscillating entities (or nodes) will sometimes lock on to a common frequency despite differences in the natural frequencies of the individual nodes. The model is relevant to a number of phenomena including ecology, biology, physics and social and organizational systems. Strogatz (2000) provides an accessible introduction and a number of surveys summarise many technical results (see Acebron *et al* (2005), Arenas *et al* (2008), Boccaletti *et al* (2006), Dorogovstev *et al* (2008)).

While Kuramoto studied the infinite complete network it is natural to consider finite networks of any topology. This would reflect particular node relationships between a subset of all node pairs, consistent with social or organizational relationships for example.

*1.1 The model*

The basic governing equation is given by

$$\dot{\theta}_i = \omega_i + k \sum_{j=1}^{n} A_{ij} \sin(\theta_j - \theta_i), \quad i = 1,...,n. \tag{1.1}$$

where $\theta_i$ are the phases of the oscillating nodes, $\omega_i$ are the natural frequencies of the nodes and $k$ is the coupling constant. Note that each $\theta_i$ is understood to denote a function of $t$. Where we wish to denote $\theta_i$ at a particular time $s$ we shall make this explicit as $\theta_i(s)$. The same applies to the derivative terms.

*1.2 Synchronisation*

It has been observed that for many initial phases (the $\theta_i(0)$) the networks *synchronize* in that the node frequencies converge to a common value with the nodes rotating at the same rate with a constant phase difference between each pair of nodes. Moreover this phenomenon appears at a critical coupling constant $k$, and as $k$ increases applies to a greater range of initial phases. Thus the network has a *frequency fixed point* characterised by

$$\dot{\theta}_i = c, \quad i = 1,...,n. \tag{1.2}$$

By summing (1.1) over all the nodes in such a case we obtain

$$nc = \sum_{i=1}^{n} \omega_i + k \sum_{i=1}^{n} \sum_{j=1}^{n} A_{ij} \sin(\theta_j - \theta_i). \tag{1.3}$$

Since $sin(-x) = -sin(x)$ the double summation in (1.3) is equal to 0. It follows that

$$c = \frac{1}{n} \sum_{i=1}^{n} \omega_i, \tag{1.4}$$

and the phase differences at the frequency fixed point satisfy,

$$c = \omega_i + k \sum_{j=1}^{n} A_{ij} \sin(\theta_j - \theta_i), \quad i = 1,...,n.$$

Since the constant $c$ equals the average phase angle for a frequency fixed point, we shall use the notation $\overline{\omega}$ to represent this.

The character, number and location of the fixed points of a network are clearly important in understanding the types of dynamics that are possible from all possible initial phases of the system (that is where the $\theta_i$ are considered in the full range $-\pi$ to $+\pi$). In particular understanding the nature of all Lyapunov stable fixed points is particularly important, since each represents a different attractor set of positive $n$-dimensional volume within the set of all states. Note that fixed points can have a range of attractor set types ranging from single points for unstable fixed points, through $m$-dimensional ($m<n$) saddle point structures for partially stable fixed points, to $n$-dimensional volumes for stable fixed points.

## 2 Stability and monotonicity

*2.1 Stability*

**Definition 2.1** Let $\{\theta_i^*, i=1,..,n\}$ be a fixed point. Thus $\theta_i^*(t) = \theta_i^*(0) + \overline{\omega} t$ and

$$\overline{\omega} = \omega_i + k \sum_{j=1}^{n} A_{ij} \sin(\theta_j^* - \theta_i^*), \quad i = 1,...,n. \tag{2.1}$$

Then $\{\theta_i^*, i=1,..,n\}$ is *asymptotically monotonic* if for some $\delta>0$ then any system $\{\theta_i, i=1,..,n\}$ with initial state

$$\theta_i(0) = \theta_i^*(0) + \delta_i \text{ where } |\delta_i| \leq \delta, \text{ and } \sum_{i=1}^{n} \delta_i = 0, \quad i = 1,..,n,$$

and subject to (1.1) converges monotonically to $\{\theta_i^*, i=1,..,n\}$ in the sense that $\theta_i$ converges monotonically to $\theta_i^*$ for each $i$.

*Remark.* The Kuramoto model operates on the differences between phase angles rather than phase angles themselves. Thus by adding a fixed constant to all the angles of a fixed point also creates a fixed point (with the same phase angle differences). For this reason to obtain convergence we consider perturbations in which the average phase angle is unaltered (see the summation condition in the above definition). This is a form of grounding condition.

*2.2 Asymptotic monotonicity*

We note that asymptotic monotonicity is a consequence of Lyapunov stability for the Kuramoto model, and this can be shown by first order approximations of the sine function near a fixed point ((see related linearization discussions of Arenas *et al* (2008), and Kalloniatis (2009)). We briefly describe this process, making the functions of t explicit. Near a fixed point $\theta_i^*$ we express $\theta_i$ as

$$\theta_i(t) = \theta_i^*(t) + \sigma_i(t)$$

where $\sigma_i(t)$ is small. We then have

$$\theta_j(t) - \theta_i(t) = \theta_j^*(t) - \theta_i^*(t) + \sigma_j(t) - \sigma_i(t)$$
$$= \theta_j^*(0) - \theta_i^*(0) + \sigma_j(t) - \sigma_i(t).$$

By using the first order approximation for sine

$$\sin(\theta_j(t) - \theta_i(t)) \approx \sin(\theta_j^*(0) - \theta_i^*(0)) + \cos(\theta_j^*(0) - \theta_i^*(0))(\sigma_j(t) - \sigma_i(t)).$$

Substituting into (1.1) gives the first order approximation

$$\dot{\sigma}_i(t) + \overline{\omega} \approx \omega_i + k \sum_{j=1}^{n} A_{ij} [\sin(\theta_j^*(0) - \theta_i^*(0)) + \cos(\theta_j^*(0) - \theta_i^*(0))(\sigma_j(t) - \sigma_i(t))], \quad i = 1,...,n.$$

This is then a first order approximation to a system of linear first order differential equations producing local solutions of the form of a sum of linear exponentials. The Lyapunov stability allows only negative exponents within the attraction region, the largest of which dominate leading to monotonic behaviour sufficiently near the fixed point. Thus a system $\theta_i$ that converges to a fixed point $\theta_i^*$ that is Lyapunov stable must have a first order approximation sufficiently near $\theta_i^*$ of the form

$$\theta_i(t) \approx \theta_i(0) + \overline{\omega} t + (\theta_i^*(0) - \theta_i(0))(1 - e^{-\lambda_i t}) \text{ for some constants } \lambda_i > 0. \quad (2.2)$$

This leads to first order approximations of the derivative of

$$\dot{\theta}_i(t) \approx \overline{\omega} + (\theta_i^*(0) - \theta_i(0))\lambda_i e^{-\lambda_i t}. \quad (2.3)$$

In particular the negative exponent means that $\theta_i(t)$ and $\dot{\theta}_i(t)$ converge in opposing senses in that $\theta_i(t)$ converges monotonically to $\theta_i^*(t)$ from above and $\dot{\theta}_i(t)$ converges monotonically to $\overline{\omega}$ from below, or $\theta_i(t)$ converges monotonically to $\theta_i^*(t)$ from below and $\dot{\theta}_i(t)$ converges monotonically to $\overline{\omega}$ from above.

## 3 Homogeneous systems

Notice that equation (2.1) can be rewritten as

$$\frac{\overline{\omega} - \omega_i}{k} = \sum_{j=1}^{n} A_{ij} \sin(\theta_j^* - \theta_i^*), \quad i = 1,...,n.$$

It follows that as the coupling constant increases in relation to the natural frequencies the left hand side tends to 0, in the limit giving the equation (3.1). Notice that this limiting equation is independent of the value of the equal frequency.

$$0 = \sum_{j=1}^{n} A_{ij} \sin(\theta_j^* - \theta_i^*), \quad i = 1,...,n. \tag{3.1}$$

Thus as the coupling constant increases, the fixed points for unequal natural frequencies (2.1) converge to the solutions for equal natural frequencies (3.1). For this reason the fixed point character of the equal frequency case is fundamental to understanding the limiting behaviour of the general case. We shall henceforth refer to the equal frequency case (leading to (3.1)) as *homogeneous* and the general case as *inhomogeneous*.

Though (3.1) can have many (indeed infinitely many) fixed points, we show for a class of networks including the complete network that there is a single stable fixed point solution, namely the zero solution $\{\theta_i^*=0, i=1,..,n\}$.

The following lemmas provide the tools we shall use to analyse stability.

## 4 The Fixed Point and Stability Lemmas

**Lemma 4.1** Let $\{\theta_i^*, i=1,..,n\}$ be any fixed point solution to the Kuramoto model. Then for any partition of the nodes into sets $A$ and $B$,

$$|A|\overline{\omega} = \sum_{i \in A} \omega_i + k \sum_{(i,j) \in (A,B)} A_{ij} \sin(\theta_j^* - \theta_i^*). \tag{4.1}$$

In particular if the frequencies are all equal we have

$$0 = \sum_{(i,j) \in (A,B)} A_{ij} \sin(\theta_j^* - \theta_i^*). \tag{4.2}$$

**Proof.** Since $\{\theta_i^*, i=1,..,n\}$ is a fixed point we have

$$\overline{\omega} = \omega_i + k \sum_{j=1}^{n} A_{ij} \sin(\theta_j^* - \theta_i^*), \quad i = 1,...,n. \tag{4.3}$$

Summing this over the nodes of A we have

$$|A|\overline{\omega} = \sum_{i \in A} \omega_i + k \sum_{i \in A} \sum_{j=1}^{n} A_{ij} \sin(\theta_j^* - \theta_i^*)$$

$$= \sum_{i \in A} \omega_i + k \sum_{(i,j) \in (A,B)} A_{ij} \sin(\theta_j^* - \theta_i^*) + k \sum_{(i,j) \in (A,A)} A_{ij} \sin(\theta_j^* - \theta_i^*).$$

Note that the second summation term is 0 since each edge is counted in both directions, which gives the result. □

**Lemma 4.2** Let $\{\theta_i^*, i=1,..,n\}$ be any stable fixed point solution to the Kuramoto model. Then for any partition of the nodes into non-empty sets $A$ and $B$,

$$\sum_{(i,j) \in (A,B)} A_{ij} \cos(\theta_i^* - \theta_j^*) > 0. \qquad (4.4)$$

Where $(A,B)$ are the set of links $(i,j)$ with $i \in A$ and $j \in B$.

*Remark.* It follows trivially from this lemma that if for every link ($A_{ij}=1$), $(\theta_i^* - \theta_j^*) > \pi/2$ then the fixed point $\{\theta_i^*, i=1,..,n\}$ is unstable. It is also well known that if for each link $(\theta_i^* - \theta_j^*) < \pi/2$ the fixed point is stable. If however there are phase differences some less than and some greater than then the fixed point may be stable or unstable (see Ochab and Gora (2009)). Lemma 4.2 provides a tool to help resolve this mixed case.

**Proof.** Since $\{\theta_i^*, i=1,..,n\}$ is a stable fixed point it is asymptotically monotonic and so monotonic within some perturbation $\delta$ (see the discussion of Section 2). Then in particular we may consider the system $\{\theta_i, i=1,..,n\}$ with initial values

$$\theta_i(0) = \theta_i^*(0) + \varepsilon |B| \quad \text{for } i \in A$$
$$\theta_i(0) = \theta_i^*(0) - \varepsilon |A| \quad \text{for } i \in B \qquad (4.5)$$

Where $\varepsilon < \delta/n$.

Since $\theta_i(0) < \theta_i^*(0)$ for $i \in A$ then by (2.3)

$$\dot{\theta}_i(0) < \overline{\omega}, \quad \text{for } i \in A. \qquad (4.6)$$

Now $\{\theta_i, i=1,..,n\}$ is subject to the original equation (1.1), so

$$\dot{\theta}_i(0) = \omega_i + k \sum_{j=1}^{n} A_{ij} \sin(\theta_j(0) - \theta_i(0)), \quad i = 1,...,n.$$

Summing this over the nodes in $A$ we have

$$\sum_{i \in A} \dot{\theta}_i(0) = \sum_{i \in A} \omega_i + k \sum_{i \in A} \sum_{j=1}^{n} A_{ij} \sin(\theta_j(0) - \theta_i(0))$$

$$= \sum_{i \in A} \omega_i + k \sum_{(i,j) \in (A,A)} A_{ij} \sin(\theta_j(0) - \theta_i(0)) + k \sum_{(i,j) \in (A,B)} A_{ij} \sin(\theta_j(0) - \theta_i(0)).$$

As in the proof of Lemma 4.1 the second summation term is equal to 0, while we can substitute from (4.5) into the last summation term, noting that $\theta_j^*(t) - \theta_i^*(t) = \theta_j^*(0) - \theta_i^*(0)$ to obtain

$$\sum_{i \in A} \dot{\theta}_i(0) = \sum_{i \in A} \omega_i + k \sum_{(i,j) \in (A,B)} A_{ij} \sin(\theta_j^* - \theta_i^* - n\varepsilon). \quad (4.7)$$

For small $\varepsilon$ we have the approximation to first order

$$\sin(\theta_j^* - \theta_i^* - n\varepsilon) \approx \sin(\theta_j^* - \theta_i^*) - n\varepsilon \cos(\theta_j^* - \theta_i^*).$$

Substituting into (4.7) we obtain

$$\sum_{i \in A} \dot{\theta}_i(0) \approx \sum_{i \in A} \omega_i + k \sum_{(i,j) \in (A,B)} A_{ij} \sin(\theta_j^* - \theta_i^*) - kn\varepsilon \sum_{(i,j) \in (A,B)} A_{ij} \cos(\theta_j^* - \theta_i^*).$$

Applying Lemma 4.1 and substituting from (4.1)

$$\sum_{i \in A} \dot{\theta}_i(0) \approx |A| \overline{\omega} - kn\varepsilon \sum_{(i,j) \in (A,B)} A_{ij} \cos(\theta_j^* - \theta_i^*).$$

This implies

$$|A| \overline{\omega} - \sum_{i \in A} \dot{\theta}_i(0) \approx kn\varepsilon \sum_{(i,j) \in (A,B)} A_{ij} \cos(\theta_j^* - \theta_i^*). \quad (4.8)$$

From (4.6) the left hand side of (4.8) is positive and the first order approximation is only possible for arbitrarily small $\varepsilon$ if the summation term is positive, which proves the lemma. □

**5 Circle diagrams**

It is useful to visualise the phase angles as points on the unit circle within the complex plane where the nodes are positioned around a circle in such a way that the phase angles are measured relative to the zero angle – here chosen to be due east of the centre, and measured in a clockwise direction. Then phase angle differences are equal to the angle subtended at the centre. We illustrate in figure 1 with circle diagrams for the homogeneous system (3.1) stable fixed points associated with the hexagonal ring, and two unstable fixed points associated with the complete network on 3 nodes and a 3-regular 8 node network.

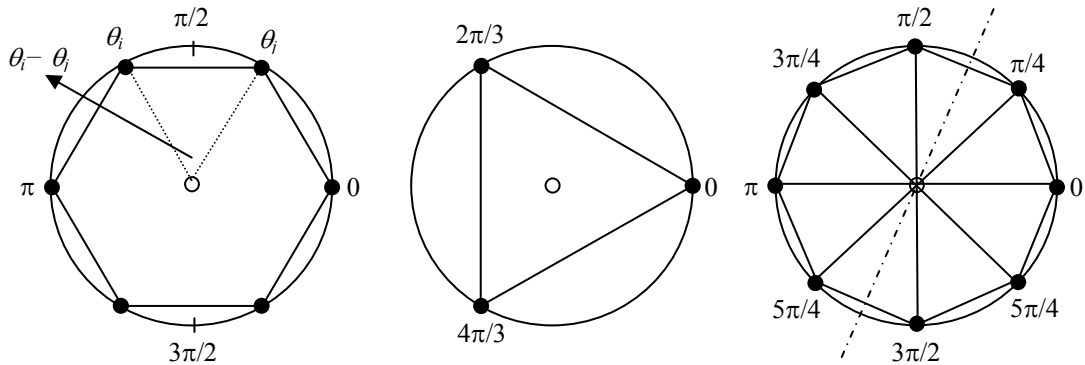

Figure 1- Fixed point networks on 6, 3, and 8 nodes.

We may use Lemma 4.2 to show that the networks on 3 and 8 vertices are unstable. For the 3 node network we consider the partition formed by taking for any node $a$, $A=\{a\}$. The left hand side of (4.4) is

$$2\cos(2\pi/3) = -1,$$

and the associated fixed point is unstable by Lemma 4.2. For the 8 node network we take $A$ to be the 4 nodes on one side of the dotted line (say at $\pi/2$, $3\pi/4$, $\pi$, and $5\pi/4$). The left hand side of (4.4) is then

$$2\cos(\pi/4) + 4\cos(\pi) = \sqrt{2} - 4 < 0,$$

and the associated fixed point is unstable by Lemma 4.2. We note that Lemma 4.2 cannot be used to show that the 6 node network of figure 1 is stable, since the Lemma is a necessary condition only.

It is well known that the zero fixed point is stable, see Arenas *et al* (2008). We show that for the homogeneous system this is the only stable fixed point for a class of networks including the complete network.

**Theorem 5.1** For any $n$ node network with degrees $n-1$ or $n-2$ the homogeneous system has no non-zero stable frequency fixed point.

**Proof.** Let $\{\theta_i^*, i=1,..,n\}$ be any non-zero fixed point solution for a network on $n$ nodes with all degrees $n$ or $n-1$.

Consider the circle diagram associated with the fixed point. First we note that the centre of the circle must lie within the outer polygon formed by the links. If this were not the case then the network would lie within one half of the circle (see figure 2) with some pair of nodes x and y with maximum phase difference.

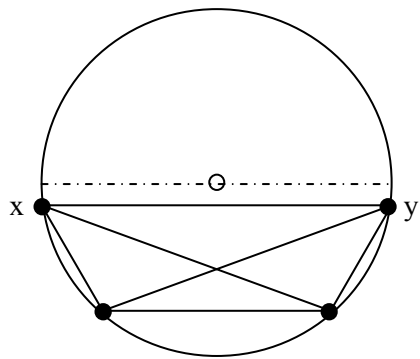

Figure 2 – Case where centre is outside the network

In this case the sine sum forming the right hand side of (4.2) at say $x$ (setting $A=\{x\}$) will have all terms of the same sign and not all zero (since the fixed point is non-zero), and so cannot sum to 0.

Represent each node as complex numbers $z_i$, $i=1,...,n$ where

$$z_j = \cos(\theta_j^*) + i\sin(\theta_j^*).$$

It then follows that (here $\bar{z}_j$ is the complex conjugate of $z_j$)

$$\operatorname{Re}\{\bar{z}_j z_k\} = \cos(\theta_j^* - \theta_k^*).$$

Let

$$S = \sum_{i=1}^{n} z_i.$$

Then

$$\operatorname{Re}\{\bar{z}_m S\} = \operatorname{Re}\left\{\bar{z}_m \sum_{i=1}^{n} z_i\right\} = \operatorname{Re}\left\{\bar{z}_m \left(z_m + \sum_{i \neq m} z_i\right)\right\}$$

$$= \operatorname{Re}\left\{\bar{z}_m z_m + \bar{z}_m \sum_{i \neq m} z_i\right\} = \operatorname{Re}\left\{1 + \sum_{i \neq m} \bar{z}_m z_i\right\}$$

$$= 1 + \sum_{i \neq m} \operatorname{Re}(\bar{z}_m z_i) = 1 + \sum_{i \neq m} \cos(\theta_i^* - \theta_m^*)$$

$$\geq \sum_{i \neq m} A_{im} \cos(\theta_i^* - \theta_m^*) \text{ since there is at most one non edge at node } m.$$

Now either $S$ is $0$ in which case the adjacency cosine sum above is less than or equal to 0 for all $m$. If $S$ is not zero then since the convex hull of the nodes contains the centre, some $z_m$ is in the opposite half plane to that centred on $S$. It follows that $\operatorname{Re}\{\bar{z}_m S\} \leq 0$ and the adjacency cosine sum is less than or equal to 0 at this $m$.

By setting $A=\{m\}$ in Lemma 4.2 this shows that the fixed point is not stable. □

It seems difficult to generalise Theorem 5.1 to include a wider class of networks, though we suspect that if all the degrees of the network are sufficiently large then there are no stable fixed points. We make two conjectures to stimulate further research. The first is suggested by the regular degree network in which the nodes are spread evenly around the circle, and are adjacent to the nearest possible neighbours. The cut-set formed by dividing the network in half is considered and Lemma 4.2 applied (see figure 3).

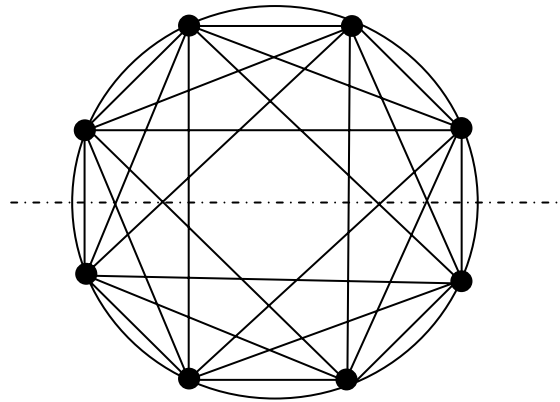

Figure 3 – regular degree network

**Conjecture 5.1** All homogeneous systems with networks on *n* nodes with all node degrees at least 2*r* can have no non-zero stable fixed point, where r is the smallest number satisfying

$$\sum_{i=1}^{r} i \cos\left(\frac{2\pi i}{n}\right) \leq 0.$$

(*2r* is approximately 0.7420*n* as *n*→∞)

**Conjecture 5.2** For the general network $A_{ij}$, determining whether or not the homogeneous system has a non-zero stable frequency fixed point is NP Hard.

**6 Conclusion**

Characterising the stable fixed points of the Kuramoto model is central to understanding the global dynamics of the model, since each stable fixed point has attractor sets of positive measure over the complete phase space. Furthermore the homogeneous Kuramoto model, apart from being of interest in its own right, represents the behaviour of the inhomogeneous model in the limit as the coupling constant tends to infinity. Thus the characterization of the stable fixed points in the homogeneous case is a valuable place to start in understanding the general (inhomogeneous) case.

We present a powerful new necessary condition for a stable fixed point, and use this to show that for the homogeneous system this is the only stable fixed point for a class of networks including the complete network (see Theorem 5.1). This result provides further evidence that in the homogeneous case the zero fixed point has an attractor set consisting of the entire space minus a set of measure zero, a conjecture of Verwoerd and Mason (2007).

Much remains to be understood however about the totality of stable fixed points in the Kuramoto model, and Theorem 5.1 seems far from being a best possible result. We therefore present two conjectures to stimulate further research.

**Acknowledgements**

The author is thankful to colleagues with whom a number of fruitful conversations have taken place over the course of this research. They are A. Dekker, M. Sweeney and A. Kalloniatis, the latter also providing detailed feedback on an early draft of this work. The argument in the latter part of Theorem 5.1, though a standard approach in complex analysis was not known to the author, and so thanks are also expressed to "Gerry" in answer to The cosine sum problem, *Mathoverflow.net*, posted August 18, 2010.